\documentclass[twocolumn,
aps,nofootinbib,showpacs,showkeys,preprint
tightenlines,preprintnumbers,
] {revtex4}

\usepackage{epsf,epsfig,subfigure,graphicx,amsmath,amssymb}
\usepackage{color}
\newcommand{\dis}[1]{\begin{equation}\begin{split}#1\end{split}\end{equation}}

\def\lsim{\lower.7ex\hbox{$\;\stackrel{\textstyle<}{\sim}\;$}}

\newcommand{\tev}{\,\textrm{TeV}}
\newcommand{\gev}{\,\textrm{GeV}}

\newcommand{\etal}{{\it et al.}}

\def\Mgl{M_{\rm gluino}}
\def\msq{m_{\rm squark}}

\def\Mmess{{M_{\rm mess}}}
\def\Mq30{{\tilde M}^{2}_{\tilde q_3}}

\def\ie{{\it i.e.}\ }

\begin{document}
%\draft
%\preprint{SNUTP 11-008 }
\title{Inverted effective SUSY with combined $Z'$ and gravity mediation, and muon anomalous magnetic moment}
%\address{Department of Physics, Tohoku University, Sendai 980-8578, Japan}
\author{Jihn E.  Kim}
\affiliation
{Department of Physics and Astronomy and Center for Theoretical
 Physics, Seoul National University, Seoul 151-747, Korea, and\\
GIST College, Gwangju Institute of Science and Technology, Gwangju 500-712, Korea}

\begin{abstract}
Effective supersymmetry(SUSY) where stop is the lightest squark may run into a two-loop tachyonic problem in some $Z'$ mediation models. In addition, a large $A$ term or/and a large stop mass are needed to have $\sim$ 126 GeV Higgs boson with three families of quarks and leptons. Thus, we suggest an inverted effective SUSY(IeffSUSY) where stop mass is larger compared to those of the first two families. In this case, it is possible to have a significant correction to the anomalous magnetic moment of muon. A three family IeffSUSY in a $Z'$ mediation scenario is explicitly studied with the $Z'$ quantum number related to $B-L$. Here, we adopt both the $Z'$ mediation and gravity mediation where the $Z'$ mediation is the dominant one for stop, while the gravity mediation is the dominant one for the muonic leptons and Higgs multiplets.
We present a numerical study based on a specific anomaly free model, and show the existence of the parameter region where all the phenomenological conditions are satisfied.
\keywords{Inverted effective SUSY, Heavy stop, $Z'$ mediation, Muon anomalous magnetic moment}
\end{abstract}

 \pacs{12.60.Jv, 14.80.Ly, 11.25.Wx, 11.25.Mj}

 \maketitle

%%%%%%%%%%%%%%%%%%%%%%%%%%%%%%%%%%%%%%%%%%%%%%%%%%%%%%%%%%%%%%%%%%%
%%%%%%%%%%%%%%%%%%%%%%%%%%%%%%%%%%%%%%%%%%%%%%%%%%%%%%%%%%%%%%%%%%%
\section{Introduction}

The recent LHC reports hint the Higgs boson mass at 125--127 GeV \cite{CERNJuly4}. This small Higgs boson mass compared to the Planck mass needs a huge hierarchy of mass scales, inviting solutions of the hierarchy problem. Supersymmetry(SUSY) has been considered to be the most attractive one among the hierarchy solutions, but the LHC data is not consistent with the constrained minimal supersymmetric standard model(CMSSM) prediction in the region $ \Mgl\msq \lesssim 1\tev^2$. A small Higgs boson mass ($m_h\simeq 0.126\,\tev$) needs a large stop mass or/and a large $A$-term in the CMSSM.

The LHC hints toward large squark masses are usually interpreted as a large mass limit for the first family squarks. The third family squarks have much lower exclusion bound \cite{LHC3rd} than those of the first two families. The current bound for $m_{\tilde q_{1,2}}$ is usually taken as 1.5\,TeV \cite{LHCsquark}. Thus, the squark masses of the third family can be below 1\,TeV in principle, which has been proposed long time ago as the effective SUSY(effSUSY) \cite{effSUSY95}.  So, if Nature has low energy SUSY as a solution to the hierarchy problem, the previously considered attractive models have been the effSUSY where only the third family squarks are in the reach of the LHC search.

Another LHC hint is the possibility of light Higgs boson whose mass is around 126 GeV \cite{CERNJuly4}. However, a light Higgs boson as heavy as 126 GeV is difficult to obtain in the CMSSM, mainly due to the tree level mass bound $m_h^{\rm tree}\le M_Z$. The loop corrections raise the Higgs mass but a fine-tuning is needed to raise it above 120 GeV \cite{HiggsBound}. This has led to scenarios with a large $A$-term from the large top Yukawa coupling and/or a large stop mass. The effSUSY through gauge mediation cannot lead to a large $A$-term since the gravity mediation for the squark mass generation is assumed to be sub-dominant compared to that of the gauge mediation. Also, the effSUSY assumes a relatively small stop mass. If gauge mediation is effectively achieved by a family-dependent $Z'$ mediation \cite{Jeong11}, then there is another problem that most scalar particles become tachyons, if  the  two-loop contributions are included.

A large hierarchy of soft masses between different families is easily realized by imposing family dependent $Z'$ charges in the $Z'$ mediation  \cite{Lang08,Zpfamind}. If the soft masses of some specific family are much smaller than those of another family, the heavy soft mass term contributes to the light soft masses at the two loop level through the SM gauge group interaction \cite{ArkaMura97}. Therefore, if the light scalars are not charged under U(1)$_{Z'}$ in the $Z'$ mediation scenario, they become tachyonic when two loop effects are taken into account.
% Recently, this issue has been studied extensively by Tamarit \cite{Tamarit12}.
This two-loop tachyonic problem may not be present in the $Z'$ D-term breaking \cite{Hisano99}. However, the model building along the D-term breaking may be more complicated than the method we introduce below with three chiral families of quarks and leptons.

The family-dependent $Z'$ mediation is so easily realized in string models \cite{Jeong11} that we consider its realization a {\it natural} one. As mentioned above, however, we need a large $A$-term and/or large stop masses from the LHC constraints.  Here, we implement the large $A$-term by the gravity mediation. Both for the $Z'$ mediation and the gravity mediation, the same dynamical SUSY breaking scale applies, which is assumed to be around $\Lambda_h\simeq 10^{13}\,\gev$ \cite{Nilles84}. Gravity mediation with this dynamical SUSY breaking scale sets the scale for the $A$-term. To radiatively raise the Higgs boson mass sufficiently above $M_Z$, we need large stop masses. So, the family-dependent $Z'$ mediation is of the form `inverted', in the sense that the 3rd family squarks are heavy compared to the first two family squarks. We call this scenario {\it inverted effective SUSY}(IeffSUSY). If stops are the heaviest sfermions, we use the term IeffSUSY irrespective of the order of the remaining sfermion masses.\footnote{Here, `inverted' is used just for the heaviest stops since effSUSY has been used for the lightest stops among sfermions \cite{effSUSY95}.}

\begin{figure}[!t]
  \begin{center}
  \begin{tabular}{c}
   {\includegraphics[width=0.35\textwidth]{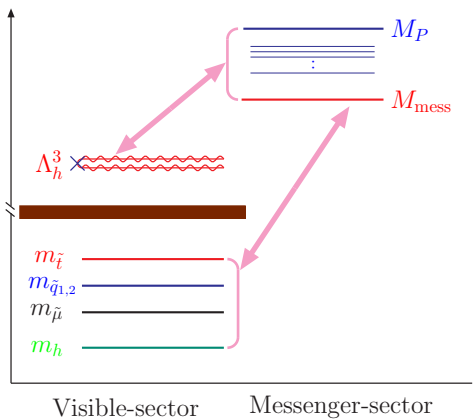}}
    \end{tabular}
  \end{center}
\caption{
{A schematic view of messenger scales and the IeffSUSY spectra. The lowest scale, assumed to be separated somewhat from the others, is called $\Mmess$.
}
   }\label{fig:Mmess}
\end{figure}

In the family-dependent $Z'$ mediation scenario,
the 3rd family members can be made heavy by assigning large $Z'$ quantum numbers to them while keeping the first two family members to carry very small $Z'$ quantum numbers, which is the opposite view taken from that of Ref. \cite{Jeong11}. %\footnote{In another context, making the 3rd family squarks heavy has been pointed out before via the top-color model in Ref. \cite{ChunLiu99}.}
So, the first two family squarks obtain masses predominantly via the gravity mediation. [Note that the 3rd family members get the additional contribution through the $Z'$ mediation.] For the gravity mediation, the messenger scale is considered to be the Planck mass $M_P=2.44\times 10^{18}\,\gev$. For the  $Z'$ mediation, the messenger scale is another parameter $\Mmess$. If $\Mmess\lesssim \frac{1}{10}M_P$, we can achieve a reasonable IeffSUSY. The messenger scales and the visible sector masses are depicted schematically in Fig. \ref{fig:Mmess}, where the visible and the hidden sectors do not communicate directly as emphasized by the thick brown bar in Fig.  \ref{fig:Mmess}.  Assuming that the lowest messenger scale $\Mmess$ is significantly separated from the other messenger scales, the low energy spectra is dominated by the scale $\Mmess$ of the $Z'$ mediation. In this sense, we argue that the $Z'$ mediation arises {\it naturally} from an ultraviolet completed theory, as far as the lowest messenger scale $\Mmess$ is sufficiently separated from the other messenger scales.

Now, the two-loop tachyons are made stable by the positive soft mass arising from the gravity mediation. In addition, in this IeffSUSY the muon $g-2$ deviation \cite{gmtwoBNL} from the SM estimation can be made significant through the light gaugino masses, which arise at the two-loop in the $Z'$ mediation \cite{Lang08}, and the light smuon ($\tilde\mu$) and scalar-muonneutrino ($\tilde\nu_2$) masses.

%%%%%%%%%%%%%%%%%%%%%%%%%%%%%%%%%%%%%%%%%%%%%%%%%%%%%%%%%%%%%%%%%%%%%%%%%%%%%%%%%%%%%%%%
%%%%%%%%%%%%%%%%%%%%%%%%%%%%%%%%%%%%%%%%%%%%%%%%%%%%%%%%%%%%%%%%%%%%%%

%%%%%%%%%%%%%%%%%%%%%%%%%%%%%%%%%%%%%%%%%%%%%%%%%%%%%%%%%%%%%%%%%%%%%%%%%%%%%%%%%%%%%%%%%%%%
\section{large correction to muon anomalous magnetic moment}\label{sec:ModelIeffSUSY}

%%%%%%%%%%%%%%%%%%%%%%%%%%%%%%%%%%%%%%%%%%%%%%%%%%%%%%%%%%%%%%%%%%%%%%%%%%%%%%%%%%%%%%%%%%%%%%
\subsubsection{Soft mass scales in the $Z'$ mediation}

As shown in Fig. \ref{fig:Mmess}, in the $Z'$ mediation scenario the soft masses at the messenger scale are generated by $Z'$ charged matter. At the messenger scale $\Mmess$, the mass splitting of $Z'$ gauge multiplet, \ie the superpartner Zprimino mass minus $Z'$ gauge boson mass, $\Delta M_{{\tilde Z}'}$, becomes
\dis{
\Delta M_{{\tilde Z}'}=-\frac{g_{Z'}^2}{8\pi^2}\frac{F}{M_{\rm mess}}=-\frac{\tilde\alpha}{2\pi}\,\frac{F}{M_{\rm mess}}
}
where $\tilde\alpha=\frac{g_{Z'}^2}{4\pi}$ and $F=\Lambda_h^3/M_P$. Then, the sfermion soft term is estimated as
\dis{
m_i^2={Z'}^2 (\Delta M_{{\tilde Z}'})^2,
}
where $Z'$ is the U(1)$_{Z'}$ charge and the visible sector gauginos $\tilde g^a$ obtain mass at the two-loop level,
\dis{
\frac{M_a(\mu)}{\alpha_a(\mu)}=-\frac{c_a\tilde\alpha (M')}{(2\pi)^2} \Delta M_{{\tilde Z}'}(M')\ln\Big(\frac{M_{\rm mess}}{M'}\Big)
}
where $M'$ is the U(1)$'$ gauge symmetry breaking scale and $c_a=\sum (Z')^2 T_s $.

The gravity mediation is taken into account such that soft terms appear at the Planck mass scale, which is imposed as the initial conditions in the following numerical study: $m_0=m_{1/2}=m_{3/2}=\Lambda_h^3/\sqrt{3}M_P^2$. This is a kind of mixed mediation studied in Refs. \cite{MixedMed2,MixedMed3}, but here again unlike in the usual gauge mediation we take a family-dependent $Z'$ mixed mediation. Hence, our results are different from those studied before  \cite{MixedMed2,MixedMed3}.

%%%%%%%%%%%%%%%%%%%%%%%%%%%%%%%%%%%%%%%%%%%%%%%%%%%%%%%%%%%%%%%%%%%%%%%%%%%%%%%%%%%%%%%%%%%%%%
\subsubsection{SUGRA model toward a large correction to $(g-2)_\mu$}

As commented in Introduction, in the CMSSM with three families of quarks and leptons it is necessary to have a large $A$-term and/or large stop mass to have a light but somewhat heavier Higgs boson above $M_Z$ at around 126 GeV. Thus, we introduce the family-dependent IeffSUSY.  If quarks and leptons in a family are treated in the simplest way, we consider a $Z'$ charge assignment related to $B-L$ as shown in Table \ref{table:BmL} where we introduce two $Z'$ mediation parameters, $\lambda_f$ for the quark and lepton superfields and $\lambda_h$ for the Higgs superfields. In our study here, we set  $\lambda_h=0$. This model does not have any gauge and gravitational anomalies.
Of course, charges may not have this simple form in string compactification.\footnote{With the spectrum of Ref. \cite{Kim07stable}, $Z'=Y+Z^{\prime\prime}$ with $Z^{\prime\prime}=-Q_1/6 -Q_4/8+ Q_5/8$ gives $Z'=0$ for $\mu^c, H_u,$ and $H_d$. With the spectrum of Ref. \cite{HuhKK09}, the heavy 3rd family and light two families and Higgs doublets, \ie $\lambda_f=\lambda_h=0$ are possible with $Z'=(0^5\, {\text -1\text  -1\, 2})(0^8)'$ in the notation of \cite{HuhKK09}.}

%%%%%%%%%%%%%%%%%%%%%%%%%%%%%%%%%%%%%%%%%%%%%%%%%%%%%%%%%%%%%%%%%
\begin{table}[!t]
\begin{center}
\begin{tabular}{|ccc|ccc|ccc|}
\hline  1$^{\rm st}$ f.  & $Y$ &$Z'$ & 2$^{\rm nd}$ f. &  $Y$ &$Z'$ &  3$^{\rm rd}$ f.  &  $Y$ &$Z'$
\\[0.2em]
\hline &&&&&&&& \\ [-1.1em]
 $(u,d)$ & $\frac16$ & $\frac{1}3\lambda_f$ & $(c,s)$ & $\frac16$ &   $\frac13 \lambda_f$ & $(t,b)$ & $\frac16$ & $\frac13 $
\\[0.4em]
$u^c $& $\frac{-2}3$ & $\frac{-1}3\lambda_f$ & $c^c$& $\frac{-2}3$ &  $\frac{-1}3 \lambda_f$ & $t^c$& $\frac{-2}3 $ &$\frac{-1}3$
\\[0.4em]
$d^c $& $\frac{1}3$ & $\frac{-1}3\lambda_f$ &$s^c$ & $\frac{1}3$ & $\frac{-1}3 \lambda_f$ &$b^c$ & $\frac{1}3$  &  $\frac{-1}3$
\\[0.4em]
 $(\nu_e,e)$& $\frac{-1}2$ & $-\lambda_f$ &$(\nu_\mu,\mu)$ & $\frac{-1}2$ & $0$&$(\nu_\tau,\tau)$& $\frac{-1}2$ & $-(1+\lambda_f)$
\\[0.4em]
$e^c$ & $1$ & $\lambda_f$ & $\mu^c$ & $1$ & $0$ & $\tau^c$& $1$ & $1+\lambda_f$
\\[0.4em]
$N^c_{1}$& $0$ & $\lambda_f$ & $N^c_{2}$ & $0$ & $0$& $N^c_{3}$& $0$ & $1+\lambda_f$
\\[0.4em]
\hline
\end{tabular}
\begin{tabular}{|ccc|}
\hline  $H_{d,u}$& $Y$ &$Z'$
\\[0.2em]
\hline  &&\\ [-1.1em]
 $H_d$ & $\frac{-1}{2}$ & $\frac{-1}{2}\lambda_h$
\\[0.4em]
 $H_u$ & $\frac{1}{2}$ & $\frac{1}{2}\lambda_h$
\\ [0.2em]
\hline
\end{tabular}
\end{center}
\caption{The $Z'$ charges of the SM fermions, Higgs doublets and heavy neutrinos. The choice $\lambda_f=\lambda_h=0$ is the simplest case. For the BNL $(g-2)_\mu$, the muonic leptons carry vanishing $Z'$ charges while the others carry nonzero $Z'$ charges with nonzero $\lambda_f$.} \label{table:BmL}
\end{table}
%%%%%%%%%%%%%%%%%%%%%%%%%%%%%%%%%%%%%%%%%%%%%%%%%%%%%%%%%%%%%%%%%%%%%%%%%%%%%%%%%%%%

Interesting cases are  $\lambda_f=\pm1$ which give the same squark masses. If these squarks are removed at very high energy scale, it is similar to but not the same as the split SUSY \cite{SplitSUSY} because scalar muons for $\lambda_f=\pm1$ and scalar taus for $\lambda_f=-1$ survive to low energy. At a first glace, these light scalar leptons may work against the gauge coupling unification since the 2nd (and the 3rd family) sfermions do not form GUT multiplets, but there is the tunable parameter $\Mmess$ for the gauge coupling unification. These particular cases of $\lambda_f$ can be called {\it partly--split SUSY}. Note also that the $\lambda_f=-1$  case allows the quantum numbers such that a large mixing between $\nu_\mu$ and $\nu_\tau$ is not forbidden.\footnote{For a sizable mixing between $\nu_e$ and $\nu_\mu$, a large $Z'=+1$ scalar VEV insertion is needed in the seesaw diagram of neutrino masses.}

%%%%%%%%%%%%%%%%%%%%%%%%%%%%%%%%%%%%%%%%%%%%%%%%%%%%%%%%%%%%%%%%%%%%%%%%%%%%%%%%%%%%%%%%%%%%%%
\subsubsection{The $\sigma_{\mu\nu}$ couplings}

The Brookhaven National Laboratory(BNL) has measured $(g-2)$ of $\mu$ with the following discrepancy on the muon anomalous magnetic moment from the SM prediction \cite{gmtwoBNL}
\dis{
\Delta a_\mu= a_\mu^{\rm exp}- a_\mu^{\rm SM}= 287(80)\times 10^{-11}\label{eq:BNLnumber}
}
where the errors including the electroweak and hadronic contributions are combined in the quadrature, resulting to the discrepancy of $3.6\,\sigma$ error.

The experimentally interesting  $\mu\to e\gamma$ mode amplitude is inversely proportional to the flavor violation quantum number $\lambda_f^2$, viz. $M_{\tilde e}^{-2}\propto \lambda_f^{-2}$. Since the present bound from $\mu\to e\gamma$ surpasses all the other lepton flavor violation bounds \cite{Chua12}, we consider only $\mu\to e\gamma$ here for the lepton flavor violation through the $\sigma_{\mu\nu}$ coupling. The $\sigma_{\mu\nu}$ coupling of neutrino was considered long time ago for the neutral current data \cite{KMO74}, which has a different chirality from the tree level $V$ and $A$ interactions of the SM. So, the $\sigma_{\mu\nu}$ couplings arise at higher orders. It must involve interactions beyond the $V-A$ charged current interactions. If it arise from the tree level $V$ and $A$ interactions, there must be $V+A$ charged current interactions also \cite{Kim76}. If we introduce the $S$ and $P$ interactions as in the SUSY extension of the SM, the $\sigma_{\mu\nu}$ couplings can be obtained at one loop level. So, the $\sigma_{\mu\nu}$ couplings are classified into a different chirality class from that of $V$ and $A$. Now, let us parametrize the one loop $\sigma_{\mu\nu}$ couplings of the charged leptons to the electromagnetic field strength $F_{\mu\nu}$ as
\dis{
{\cal L}= A_{L'R}\, \bar l\,'_L \sigma_{\mu\nu} &l_R F^{\mu\nu} + A_{R'L}\, \bar l\,'_R \sigma_{\mu\nu} l_L F^{\mu\nu}\\
&(+{\rm h.c.~if~}l\ne l').
}
Note that the real parts of the flavor diagonal $A_{\mu_L \mu_R}$ and $A_{\mu_R \mu_L}$ contribute to  $(g-2)_\mu$, the imaginary parts of $A_{\mu_L \mu_R}$ and $A_{\mu_R \mu_L}$ contribute to the muon EDM, and the flavor violating absolute magnitude contributes to   BR$(\mu\to e\gamma)$. Parametrizing $g-2$, EDM and $\mu\to e\gamma$ Lagrangians with $\sigma_{\mu\nu}$ couplings as \cite{Chua12,Moroi96}
\dis{
&{\cal L}_{g-2} =\frac{e}{2m_\mu}\Delta a_\mu \, \bar\mu\,\sigma_{\mu\nu} \mu F^{\mu\nu}  ,\\
&{\cal L}_{EDM} =-\frac{i e}{2} \delta_\mu  \, \bar\mu\,\sigma_{\mu\nu} \gamma_5 \mu F^{\mu\nu},\\
&{\cal L}_{\mu\to e\gamma} =\frac{e}{4m_\mu}\epsilon_{e \mu} \, \bar e\,\sigma_{\mu\nu} \mu F^{\mu\nu}  +{\rm h.c.},
}
where the $\mu\to e\gamma$ Lagrangian is compared to the $(g-2)_\mu$ Lagrangian with the flavor violating parameter $\epsilon_{e \mu}$, and we may factor out $e$ from the muon EDM $d_\mu$: $d_\mu\equiv \delta_\mu e$. Then, we have a chirality relation on the coefficients of $\sigma_{\mu\nu}$ couplings,
\dis{
4\,\frac{\Gamma(\mu\to e\gamma)}{\epsilon_{e \mu}^2 m_\mu}=  (\Delta a_\mu)^2 + \delta_\mu^2 .
}

The BR($\mu\to e\gamma$) is given by \cite{Chua12},
\dis{
{\rm BR}(\mu\to e\gamma) &=\frac{48\pi^2}{G_F^2 m_\mu^2}\left(|A_{e_L \mu_R}|^2 +|A_{e_R\mu_L}|^2\right)\\
&=\frac{24\pi^3\alpha_{\rm em}|\epsilon_{e\mu}|^2}{G_F^2 m_\mu^4}=3.2 \times 10^{14} |\epsilon_{e\mu}|^2 .
}
If selectron is heavier than Zino, we estimate $\epsilon_{e\mu}\sim g^2 m_\mu M_{\tilde Z}\Delta M_{\tilde e\,\tilde\mu}^2 /32\pi^2 M_{\tilde \mu}^2 M_{\tilde e}^2$,  and obtain
\dis{
|\epsilon_{e\mu}|^2 \sim \frac{ \alpha_{\rm em}^2 m_\mu^2 M_{\tilde Z}^2  (\Delta M_{\tilde e\,\tilde\mu}^2)^2}{64 \pi^2\sin^4\theta_W M_{\tilde \mu}^4 M_{\tilde e}^4 }.
}
Therefore, BR($\mu\to e\gamma$) is estimated as
\dis{
{\rm BR}(\mu\to e\gamma) \sim 5.7\times 10^{2}\left(\frac{M_{\tilde Z,\,100}^2}{ M_{\tilde \mu,\,100}^4} \right)\left(\frac{\Delta M_{\tilde e\,\tilde\mu}^2}{ M_{\tilde e}^2 }
\right)^2
}
where $\Delta M_{\tilde e\,\tilde\mu}^2$ is the parameter denoting the slepton mixing, and $M_{\tilde Z,\,100}$ and $M_{\tilde \mu,\,100}$ are $M_{\tilde Z}$ and $M_{\tilde \mu}$ masses in units of 100\,GeV. This can be compared to the experimental upper bound $2.4\times 10^{-12}$ \cite{Adam11}.

%%%%%%%%%%%%%%%%%%%%%%%%%%%%%%%%%%%%%%%%%%%%%%%%%%%%%%%%%%%%%%%%%%%%%%%%%%%%%%%%%%%%%%%%
\begin{figure}[!t]
  \begin{center}
  \begin{tabular}{c}
   {\includegraphics[width=0.32\textwidth]{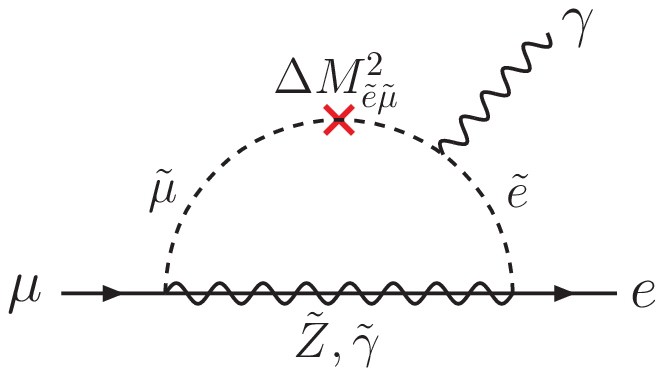}}
    \end{tabular}
  \end{center}
\caption{
{A $\mu\to e\gamma$ diagram.
}
   }\label{fig:MixMuEl}
\end{figure}
%%%%%%%%%%%%%%%%%%%%%%%%%%%%%%%%%%%%%%%%%%%%%%%%%%%%%%%%%%%%%%%%%%%%%%

A typical flavor mixing diagram is shown in Fig. \ref{fig:MixMuEl}.
If the flavor mixing parameter $\Delta M_{\tilde e\,\tilde\mu}^2$ is assumed to be $M_{\tilde \mu}^4/M^2_{\tilde e}$, the RHS is proportional to $(M_{\tilde \mu}/M_{\tilde e})^8$ and the selectron mass more than
60 times the smuon mass is enough to satisfy the $\mu\to e\gamma$ upper bound, assuming the order 1 value for $M_{\tilde Z,\,100}^2/ M_{\tilde \mu,\,100}^4$. If $\lambda_f$ is large, selectron mass is few tens times the smuon mass. If $\lambda_f$ is small, e.g. for $\lambda_f=\frac16$ the ratio is about 4. But the flavor problem can be accurately addressed in a complete model with the details of the first two family Yukawa couplings, which we do not consider in this paper.

%%%%%%%%%%%%%%%%%%%%%%%%%%%%%%%%%%%%%%%%%%%%%%%%%%%%%%%%%%%%%%%%%%%%%%%%%%%%%%%%%%%%%%%%%%%%%%
\subsubsection{Muon anomalous magnetic moment from SUGRA chargino and neutralino}

%%%%%%%%%%%%%%%%%%%%%%%%%%%%%%%%%%%%%%%%%%%%%%%%%%%%%%%%%%%%%%%%%%%%%%%%%%%%%%%%%%%%%%%%
\begin{figure}[!t]
  \begin{center}
  \begin{tabular}{c}
   {\includegraphics[width=0.32\textwidth]{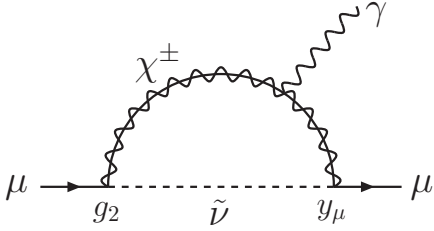}}
    \end{tabular}
  \end{center}
\caption{
{A chargino diagram toward $(g-2)_\mu$.
}
   }\label{fig:gminustwo}
\end{figure}
%%%%%%%%%%%%%%%%%%%%%%%%%%%%%%%%%%%%%%%%%%%%%%%%%%%%%%%%%%%%%%%%%%%%%%

The muon anomalous magnetic moment in the MSSM has been given in Ref. \cite{Moroi96}. The anomalous magnetic moment arises from the SUSY diagrams among which the chargino diagram is shown in Fig. \ref{fig:gminustwo}. We discuss the chargino diagram here in detail in Appendix A because the results can be presented as closed forms. The neutralino diagram also contributes and the approximate form is present in Appendix B. In the mass region of our interest, it is estimated that the contribution of the neutralino diagram is more important compared to that of the chargino diagram, unlike the comments of \cite{Moroi96,Masiero07}. In the numerical analyses of Sec. \ref{sec:DataFit}, we  include both contributions. The anomalous magnetic moment $a^{\rm SUSY}_\mu$ in units of the muon Bohr magneton is the sum of mass eigenstates chargino and neutralino contributions:
\dis{
&a^{\rm SUSY}_\mu \simeq a^{\rm SUSY}_\mu (\chi^+_1)+a^{\rm SUSY}_\mu (\chi^+_2)\\
&~~~~~~~~~~+a^{\rm SUSY}_\mu (\chi^{0A}_1)+a^{\rm SUSY}_\mu (\chi^{0A}_2),\\[0.3em]
&a^{\rm SUSY}_\mu (\chi^+_1)\simeq -\frac{3m_\mu}{16\pi^2}C_X^L C_X^R \frac{m_{\chi^+_1} }{m^2_{\tilde\nu}} I^+ (x_1) ,\\
&a^{\rm SUSY}_\mu (\chi^+_2)\simeq -\frac{3m_\mu}{16\pi^2}C_X^L C_X^R \frac{m_{\chi^+_2} }{m^2_{\tilde\nu}} I^+ (x_2) ,\\
&a^{\rm SUSY}_\mu (\tilde\mu_{A1})\simeq -\frac{m_\mu}{16\pi^2}N_{AX}^L  N_{AX}^R \frac{m_{\chi^0_1} }{m^2_{\tilde\mu_{A1}}} I^0 (x_{A1}) ,\\
&a^{\rm SUSY}_\mu (\tilde\mu_{A2})\simeq -\frac{m_\mu}{16\pi^2} N_{AX}^LN_{AX}^R \frac{m_{\chi^0_1} }{m^2_{\tilde\mu_{A2}}} I^0 (x_{A2}) ,\label{eq:anommom}
}
where $I^+(x)=(1-\frac43 x +\frac13 x^2  +\frac23\ln x ) / (1-x )^3,\, I^0(x)=(1-x^2+2x\ln x)/(1-x)^3$,
$x_{1,2}\equiv m^2_{\chi_{1,2}}/m^2_{\tilde\nu_2},\, x_{A1,A2}\equiv m^2_{\chi^{0A}}/m^2_{\tilde\mu_{1,2}}$, and we calculate $C_X^L C_X^R$ and $ N_{AX}^LN_{AX}^R $ in Appendices A and B, respectively. Note that  $C_X^L= y_\mu (U_{\chi^-})_{2X}$ and $C_X^R=-g_2 (U_{\chi^+})_{1X}.$
Thus, we obtain
\dis{
\sum_X C_X^L C_X^R= y_\mu g_2(-\sin\epsilon \cos\epsilon' +\cos\epsilon \sin\epsilon').
}
Here, $C_X^L C_X^R$ contains a couplings factor $y_\mu g_2\simeq 6.05\times 10^{-4}\sqrt{1+\tan^2\beta}\times 0.6521=3.95\times 10^{-4}\sqrt{1+\tan^2\beta}$.
Thus, Eq. (\ref{eq:anommom}) becomes
\begin{widetext}
\dis{
\frac{a^{\rm SUSY}_\mu (\chi_{1,2})}{1\times 10^{-9}}\simeq
&-2.64\sqrt{1+\tan^2\beta} \,(\cos\epsilon \sin\epsilon' -\sin\epsilon \cos\epsilon' )  \left(\frac{300\,\gev}{m_{\tilde\nu_2}} \right)\left(\frac{m_{\chi_{1,2}}}{m_{\tilde\nu_2}} \right)\cdot I^+(x_{1,2})\\
&-1.32\, \tan\beta \sum_{i=\rm lighter~one}\, \left(\frac{\mu}{10\,\tev} \right)\left(\frac{(300\,\gev)^2}{|M^2_{LRZ}|} \right)
\left(\frac{\tev}{m_{\tilde\mu_{i}}} \right)^2 \left(\frac{m_{\chi^{0A}_I}}{\tev} \right)\cdot I^0(x_{Ai})
\label{eq:KimCalc}
}
%\end{widetext}
Note that the four possible cases of the chargino contribution give the same sign for the combinations $\cos\epsilon \sin\epsilon'$ and $-\sin\epsilon \cos\epsilon' $, which are of order $M_W/\mu$. From Eq. (\ref{eq:KimCalc}), we note that the neutralino contribution dominates for a large $\mu$.
In our presentation in Sec. \ref{sec:DataFit}, we work for two almost degenerate neutralinos. This degeneracy may be violated by the running of $g_2$ and $g_1$ gauge couplings, but our objective on the order estimation will not be changed much even if the running effects are taken into account.
Looking at the front numerical factor of Eq. (\ref{eq:KimCalc}), we note that the BNL $(g-2)_\mu$ can be explained by the TeV scale SUSY parameters as we show in Sect. \ref{sec:DataFit}. This scenario can be  achieved with the IeffSUSY $Z'$ quantum numbers shown in Table \ref{table:BmL}, where muonic leptons carry the vanishing $Z'$ quantum number such that their scalar partners obtain SUSY breaking masses only through gravity mediation which is of order $m_{3/2}$. These muonic scalar partners are the lightest sfermions. The next order sfermion masses are the first family sfermions and the second family squarks. The heaviest sfermions are the third family members.

%\begin{widetext}

%%%%%%%%%%%%%%%%%%%%%%%%%%%%%%%%%%%%%%%%%%%%
\section{Phenomenological bounds}\label{sec:DataFit}

%%%%%%%%%%%%%%%%%%%%%%%%%%%%%%%%%%%%%%%%%%%%%%%%%%%%%%%%%%%%%%%%%%%%%%%%%%%%%%%%%%%%%%%%
\begin{figure}[!b]
  \begin{center}
  \begin{tabular}{l}
  {\includegraphics[width=0.9\textwidth]{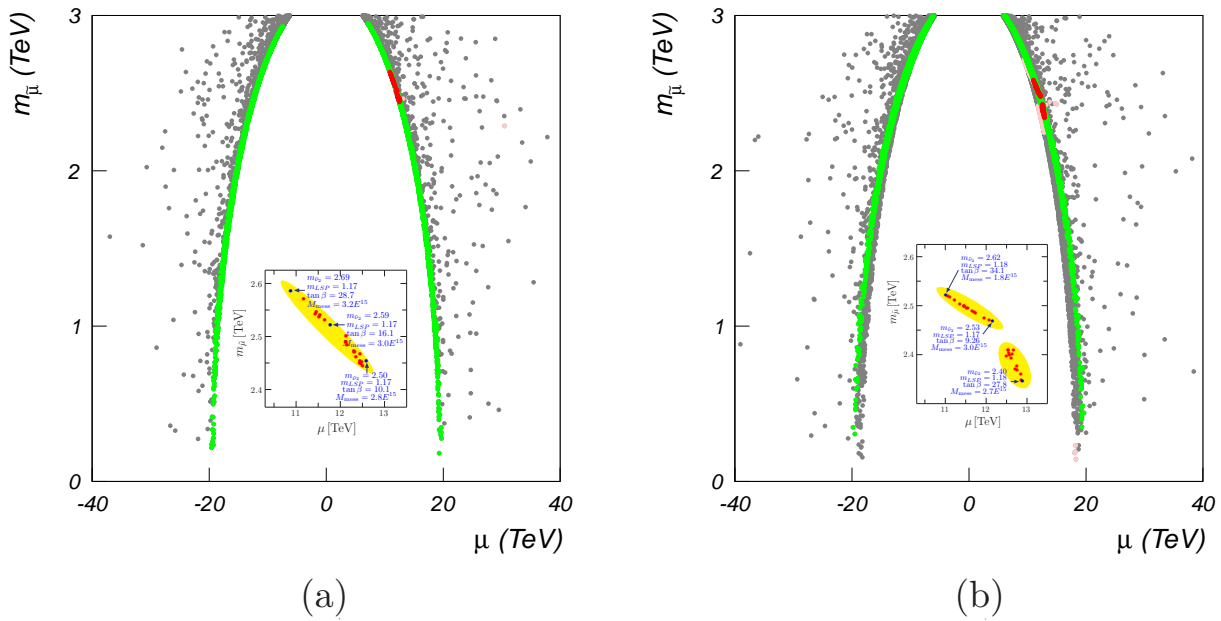}}
    \end{tabular}
\caption{
The scatter plot in the $m_{\tilde\mu}-\mu$ space for $\Lambda_h=3\times 10^{13}\gev$ out of $10^5$ trial points: (a) for  $\lambda_f=\frac16,\,\lambda_h=0$, and (b) for  $\lambda_f=-1,\,\lambda_h=0$. The scanned parameters are $\mu$ and $\Mmess=(0.1\sim 10)\times 10^{15}\,\gev$. The top quark mass corresponds to $m_t=(173.5\pm 1.4)\,\gev$. The gray dots are the trial points. The green dots are those satisfying the LHC constraints ($m_{\tilde q_{1,2}}>1.5\,\tev$ and the LHC gluino mass bound) and $m_h=(125\sim 127)\,\gev$. The pink dots are filtered by $(g-2)_\mu$. The red dots are those satisfying all the constraints including  $(g-2)_\mu$. In the enlarged insets, some selected red points are shown again with more information in blue dots .
}\label{fig:ConstraintsSmu}
  \end{center}
\end{figure}
%%%%%%%%%%%%%%%%%%%%%%%%%%%%%%%%%%%%%%%%%%%%%%%%%%%%%%%%%%%%%%%%%%%%%%

For Table \ref{table:BmL}, let the mass scale of the heavy 3rd family squarks is universally given as $\Mq30$. Assuming only $\tilde t$ and $\tilde b$ masses get $Z'$ mediated heavy mass, the two loop running equation for squarks ($i=1,2,3$), in the limit $m^2_{\tilde t} (\mu) \gg  m^2_{\tilde q_{1,2}} (\mu) $, is
\dis{
\Lambda \frac{d m_i^2}{d\Lambda} =\frac{\alpha_3^2(\Lambda)}{3\pi^2}m^2_{\tilde t} (\Lambda) \label{eq:heavytbcalar}
}
\end{widetext}
where $m^2_{\tilde b}=m^2_{\tilde t}$ has been assumed. But, the phenomenological requirements we impose is not as strong as $m^2_{\tilde t} (\mu) \gg  m^2_{\tilde q_{1,2}} (\mu)$ but simply
\dis{
&m_{\tilde q_{1,2}} >1.5\, \tev,\\
   & m_{\tilde t} = {\rm large~ enough~ to~ give~ 126\, \gev~ Higgs}.\label{eq:sqmassBounds}
}
These conditions are much more stronger than the absence of tachyons.
Therefore,  we use the package SOFTSUSY \cite{SOFTSUSY12} which includes two-loop evolutions. Given the boundary values of the input parameters such as  $m^2_{\tilde b}=m^2_{\tilde t}$ at $\Mmess$ and $m_{3/2}$ at $M_P$, the package SOFTSUSY gives the flavor dependent mass spectrum, $\tan\beta$ and $m_h$ at the TeV scale.

In our model, there are three parameters: $\Lambda_h, \Mmess,$ and $\mu$. Parameters $m_{3/2}$ and $A_t$ are given by $\Lambda^3/\sqrt{3}M_P^2$ and $f_t m_{3/2}$ (with $f_t$ chosen within $1\sigma$ at the top mass region), respectively, and $\tan\beta$ is calculated through $v_u/v_d$ by solving the running equations of soft terms $m^2_{H_u}$  and $m^2_{H_d}$, which are included in the package SOFTSUSY.

The main constraints we use are: (1) Eq. (\ref{eq:sqmassBounds}) which is stronger than no tachyon constraints of \cite{ArkaMura97}, (2) the LHC bounds on the gluino mass and the first two family squark masses \cite{LHC3rd, LHCsquark}, and  (3) the $1\sigma$ allowed region for the BNL $(g-2)_\mu$ \cite{gmtwoBNL}. The $b\to s\gamma$ constraint \cite{bsgamma} is satisfied since we required $m_{\tilde q_{1,2}}>1.5\,\tev$. In Fig. \ref{fig:ConstraintsSmu}, we present the scatter plots  in the $m_{\tilde\mu}$ vs. $\mu$ for the fixed $\Lambda_h=3\times 10^{13}\,\gev$ and the scattered $\mu$ and $\Mmess$, in (a) for  $\lambda_f=\frac16,\,\lambda_h=0$ and in (b) for  $\lambda_f=-1,\,\lambda_h=0$. The light Higgs scalar are tuned to 125$\sim$127\,GeV. The green dots of Fig. \ref{fig:ConstraintsSmu} give $m_h$ in the region 125-127 GeV, and pink dots give the BNL $(g-2)_\mu$. The red dots are satisfying all the constraints. In the insets, we present some selected red dots with more information in blue dots. For three blue dots in each inset we show the detail information on scalar-muon-neutrino mass, the LSP mass $m_{LSP}$, $\tan\beta$, and the messenger scale. We note that the LSP mass does not change very much among the blue dots (and hence red dots also).
The lightest SUSY particle is a mixture of wino and Higgsino with the mass little bit above 1\,TeV as indicated for three blue dots.

%%%%%%%%%%%%%%%%%%%%%%%%%%%%%%%%%%%%%%%%%%%%%%%%%%%%%%%%%%%%%%%%%%%%%%%%%%%%%%%%%%%%%%%%
\begin{figure}[!t]
  \begin{center}
  \begin{tabular}{c}
   {\includegraphics[width=0.45\textwidth]{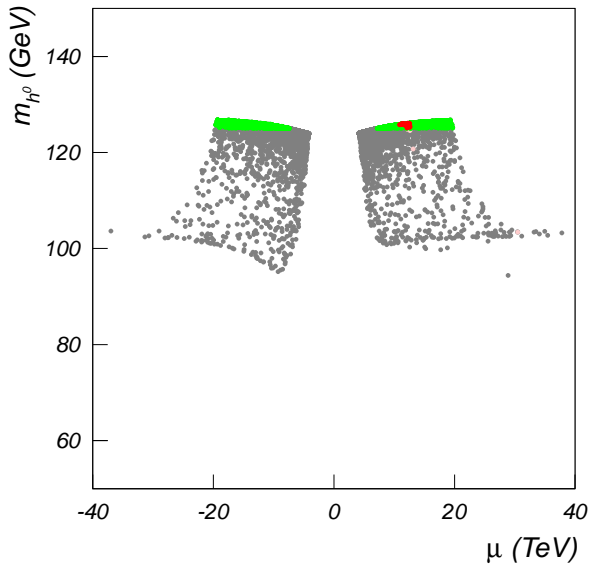}}
    \end{tabular}
  \end{center}
\caption{ Same as Fig. \ref{fig:ConstraintsSmu}\,(a) except for $m_h$ instead of $m_{\tilde\mu}$.
   }\label{fig:ConstraintsMh}
\end{figure}
%%%%%%%%%%%%%%%%%%%%%%%%%%%%%%%%%%%%%%%%%%%%%%%%%%%%%%%%%%%%%%%%%%%%%%

In Fig. \ref{fig:ConstraintsMh}, we present a scatter plot of the Higgs boson mass for the case of Fig. \ref{fig:ConstraintsSmu}\,(a).

A TeV-scale LSP in the MSSM with the R-parity conservation overcloses the universe. This overclosure problem can be evaded if the R-parity is not exact or there is a singlet heavy particle(s) decaying to the LSP as in the case of the heavy axino \cite{AxinoToLSP}.

%%%%%%%%%%%%%%%%%%%%%%%%%%%%%%%%%%%%%%%%%%%%
\section{Conclusion}\label{sec:Conclusion}

In view of the observed 126 GeV Higgs boson \cite{CERNJuly4} which is relatively heavy in the SUSY scenario, we introduced a heavy stop scheme which is the opposite view from the popular effSUSY idea. In the $Z'$ mediation scenario, we achieve this IeffSUSY explicitly with the U(1)$_{Z'}$ quantum numbers shown in Table \ref{table:BmL}. With the quantum numbers of Table \ref{table:BmL},
it is possible to have relatively light smuons ($\sim 2-3\,\tev$) and neutralino ($\sim 1.2\,\tev$),
and hence can find a parameter region where a significant correction to the anomalous magnetic moment of muon can result.

%%%%%%%%%%%%%%%%%%%%%%%%%%%%%%%%%%%%%%%%%%%%%%%%%%%%%%%%%%%%%%%%%%%%%%%%%%%%
\section*{Appendix A}

Let us express the chargino mass matrix as $-{\cal L} =\tilde\psi^{+T} M_{\chi^\pm}\tilde\psi^-$ where \cite{Moroi96}
\dis{
M_{\chi^\pm}=\left(\begin{array}{cc}
     -m_{G2} & \sqrt{2}M_W \cos\beta\\
     -\sqrt{2}M_W \sin\beta & \mu
\end{array}\right),\label{eq:MatrixChargino}
}
and
\dis{
\tilde\psi^+ &=\left( \begin{array}{c}  -i\tilde W^+\\ {\tilde H}^+_u
\end{array}\right),~
\tilde\psi^-=\left( \begin{array}{c}  -i\tilde W^-\\ {\tilde H}^-_d
\end{array}\right). \nonumber
}
The mass eigenvalues are
\dis{
&m_{\chi_1}= \sqrt{\frac{\mu^2+m_{G2}^2+2M_W^2-\sqrt{D}}{2}}\\
&m_{\chi_2}= \sqrt{\frac{\mu^2+m_{G2}^2+2M_W^2+\sqrt{D}}{2}}
}
where
\dis{
D &=\mu^4+m^4_{G2}+4M_W^4-2\mu^2 m^2_{G2} +4\mu^2 M_W^2 \\
&+4 m^2_{G2} M_W^2-16M_W^4 c_\beta^2 s_\beta^2 +16\mu  m_{G2} M_W^2 c_\beta s_\beta
}
where $c_\beta=\cos\beta$ and $s_\beta=\sin\beta$.
For the chargino diagram of Fig. \ref{fig:gminustwo}, we obtain the closed forms for the mass diagonalizing unitary matrices. The anomalous magnetic moment $a^{\rm SUSY}_\mu(\chi_i^+)$ contains these matrix elements $C_X^L\,(I=\{A,B\})$ and  $C_X^R\,(J=\{A',B'\})$ which are calculated as
\dis{
(A)\left\{\begin{array}{l}
\cos\epsilon= \left[\frac12\left(1\mp \sqrt{E } \right)\right]^{1/2}\\
\sin\epsilon= \left[\frac12\left(1\pm \sqrt{E } \right)\right]^{1/2}
\end{array}\right.\label{eq:CaseA}
}
\dis{
(B)\left\{\begin{array}{l}
\cos\epsilon= -\left[\frac12\left(1\mp \sqrt{E } \right)\right]^{1/2}\\
\sin\epsilon= -\left[\frac12\left(1\pm \sqrt{E } \right)\right]^{1/2}
\end{array}\right.\label{eq:CaseB}
}
\dis{
(A')\left\{\begin{array}{l}
\cos\epsilon'= -\left[\frac12\left(1\mp \sqrt{F} \right)\right]^{1/2}\\
\sin\epsilon'= \left[\frac12\left(1\pm \sqrt{F } \right)\right]^{1/2}
\end{array}\right.\label{eq:CaseApr}
}
\dis{
(B')\left\{\begin{array}{l}
\cos\epsilon'= \left[\frac12\left(1\mp \sqrt{F } \right)\right]^{1/2}\\
\sin\epsilon'= -\left[\frac12\left(1\pm \sqrt{F } \right)\right]^{1/2}
\end{array}\right.\label{eq:CaseBpr}
}
where $E=1-8M_W^2\cos^2\beta(\mu+m_{G2}\tan\beta)^2/D,\, F=1-8M_W^2\cos^2\beta(\mu\tan\beta+m_{G2})^2/D , \tan\beta=v_u/v_d$, and $m_{G2}$ is defined in Eq. (\ref{eq:MatrixChargino}). For $C_X^L C_X^R$, we consider four cases of $(IJ)$, where $I=\{A,B\}$ and $J=\{A',B'\}$. Among these $(AA')$ and $(BB')$ give  positive sign and  $(AB')$ and $(BA')$ give  negative sign.

\vskip 0.5cm

%%%%%%%%%%%%%%%%%%%%%%%%%%%%%%%%%%%%%%%%%%%%%%%%%%%%%%%%%%%%%%%%%%%%%%%%%%%%
\section*{Appendix B}

We diagonalize the neutralino mass matrix $M_{\chi^0}$,
\dis{
\left(\begin{array}{cccc}
-m_{G1}, & 0, & -M_Zs_W c_\beta, & M_Z s_W s_\beta \\
 0,&-m_{G2},  & M_Z c_W c_\beta , &-M_Z c_W s_\beta \\
 -M_Z s_W c_\beta , & M_Z c_W c_\beta , & 0 & \mu\\
M_Z s_W s_\beta , & -M_Z c_W s_\beta ,   & \mu & 0
\end{array}\right)
}
where $c_W=\cos\theta_W,~s_W=\sin\theta_W ,~c_\beta=\cos\beta,~s_\beta=\sin\beta $.
By the biunitary transformation by $U^\dagger_{\chi^0}$ on the left and $U_{\chi^0}$ on the right, it is diagonalized
\dis{
U^\dagger_{\chi^0}M_{\chi^0}  U_{\chi^0}  ={\rm Diag.\,}(\lambda_1,\,\lambda_2,\, \lambda_3,\,\lambda_4)
}
where we ordered $|\lambda_i|\le |\lambda_j|$ if $i<j$. Here, $U_{\chi^0}$ is a $4\times 4$ orthogonal matrix since there is no phase. Defining
 \dis{
U_{\chi^0}\equiv \left(\begin{array}{cccc}
     1 & 0& 0& 0 \\
 0 & 1 & 0& 0\\
    0 & 0& \frac{1}{\sqrt2}& \frac{1}{\sqrt2} \\
    0 & 0& \frac{1}{\sqrt2}& -\frac{1}{\sqrt2}
\end{array}\right) \tilde U,
}
we obtain

\begin{widetext}

{\tiny
\dis{
 &\left(\begin{array}{cccc}
-m_{G1}, & 0, & \frac{M_Z}{\sqrt2}s_W(- c_\beta +s_\beta), & -\frac{M_Z}{\sqrt2}s_W (c_\beta+s_\beta) \\
 0,&-m_{G2},  & \frac{M_Z}{\sqrt2}c_W (c_\beta-s_\beta), &\frac{M_Z}{\sqrt2}c_W (c_\beta+s_\beta)\\
 \frac{M_Z}{\sqrt2}s_W(- c_\beta +s_\beta), &\frac{M_Z}{\sqrt2}c_W (c_\beta-s_\beta), & \mu & 0\\
-\frac{M_Z}{\sqrt2}s_W (c_\beta+s_\beta) , & \frac{M_Z}{\sqrt2}c_W (c_\beta+s_\beta),   & 0 & -\mu
\end{array}\right)    =\tilde U\left(\begin{array}{cccc}
     \lambda_1 & 0 & 0& 0\\
     0 & \lambda_2& 0& 0\\
     0 & 0&\lambda_3& 0\\
     0 & 0& 0&\lambda_4\\
\end{array}\right)\tilde U^T.\label{eq:MassDiagDef}
}
}
In the limit $\frac{M_Z}{\mu}\simeq 0$ and  $\frac{M_Z}{m_{Gi}}\simeq 0$, the mass matrix is almost diagonalized, with the eigenvalues $\lambda_1\simeq m_{G_1}, \lambda_2\simeq m_{G_2},\lambda_3=-\lambda_4\simeq \mu$. Keeping the small parameters, we notice that there seems to be four small parameters from Eq. (\ref{eq:MassDiagDef}), $\frac{M_Z}{\sqrt2 \mu}c_W (c_\beta+s_\beta),\frac{M_Z}{\sqrt2 \mu}c_W (c_\beta-s_\beta),\frac{M_Z}{\sqrt2 \mu}s_W (c_\beta+s_\beta),$ and $\frac{M_Z}{\sqrt2 \mu}s_W (c_\beta-s_\beta)$.  A general $4\times 4$ orthogonal matrix has six real parameters. In view of the mass hierarchy $M_Z\ll m_{Gi}\ll\mu$, two large parameters and four small parameters can be a good approximate description of the $4\times 4$ orthogonal matrix. Thus, parametrizing $\tilde U$ approximately as
\dis{
\tilde U &=\left(\begin{array}{cccc}
     c_M-\frac{1}{2c_M}(\epsilon_1^2+\epsilon_4^2) & s_M  & \epsilon_1& \epsilon_4\\
      -s_M &  c_M-\frac{1}{2c_M}(\epsilon_2^2+\epsilon_3^2)  & \epsilon_3& \epsilon_2\\
     -\epsilon_1 & -\epsilon_3&c_\mu-\frac{1}{2c_\mu}(\epsilon_1^2+\epsilon_3^2)  &  s_\mu\\
     -\epsilon_4& -\epsilon_2&  -s_\mu& c_\mu-\frac{1}{2c_\mu}(\epsilon_2^2+\epsilon_4^2)
\end{array}\right),
}
where the simplified trigonometric notations $s_i=\sin\theta_i, c_i=\cos\theta_i,$ and $ t_i=\tan\theta_i$ are used. Then, the RHS of (\ref{eq:MassDiagDef}) becomes
{\tiny
\dis{
\left(\begin{array}{cccc}
\begin{array}{c} \lambda_1+\\
+ (\lambda_4-\lambda_1)(\epsilon_1^2+\epsilon_4^2)\\
+O(\epsilon_i^4)
\end{array},
& \begin{array}{c}   \\
+\frac{\lambda_1}{2}t_M(\epsilon_1^2+\epsilon_4^2\\
-\epsilon_2^2-\epsilon_3^2)\\
+\lambda_4(\epsilon_1\epsilon_3+ \epsilon_2\epsilon_4)
\end{array},
&  \begin{array}{c} -\lambda_1 (c_M\epsilon_1+ s_M\epsilon_3) \\
  +\lambda_4( c_\mu\epsilon_1+  s_\mu\epsilon_4)\\
  +O(\epsilon_i^3)
\end{array},
& \begin{array}{c} -\lambda_1 (c_M\epsilon_4+ s_M\epsilon_2)\\
  +\lambda_4(- s_\mu\epsilon_1+  c_\mu\epsilon_4)\\
  +O(\epsilon_i^3)
\end{array}\\[0.5em]
\begin{array}{c}   \\
+\frac{\lambda_1}{2} t_M(\epsilon_1^2+\epsilon_4^2\\
-\epsilon_2^2-\epsilon_3^2)\\
+\lambda_4(\epsilon_1\epsilon_3+ \epsilon_2\epsilon_4)
\end{array},
 & \begin{array}{c} \lambda_1+\\
  +(\lambda_4-\lambda_1)(\epsilon_2^2+ \epsilon_3^2)\\
+O(\epsilon_i^4)
\end{array},
& \begin{array}{c} \lambda_1 (s_M\epsilon_1-  c_M\epsilon_3) \\
  +\lambda_4( c_\mu\epsilon_3+  s_\mu\epsilon_2)\\
  +O(\epsilon_i^3)
\end{array},
& \begin{array}{c} \lambda_1 (s_M\epsilon_4- c_M\epsilon_2) \\
  + \lambda_4(- s_\mu\epsilon_3+ c_\mu\epsilon_2)\\
  +O(\epsilon_i^3)
\end{array}\\[0.5em]
\begin{array}{c} -\lambda_1 (c_M\epsilon_1+ s_M\epsilon_3) \\
  +\lambda_4( c_\mu\epsilon_1+  s_\mu\epsilon_4)\\
  +O(\epsilon_i^3)
\end{array},
 & \begin{array}{c} \lambda_1 (s_M\epsilon_1- c_M\epsilon_3) \\
  +\lambda_4( c_\mu\epsilon_3+ s_\mu\epsilon_2)\\
  +O(\epsilon_i^3)
\end{array}
 &\begin{array}{c} \lambda_4+  \\
-(\lambda_4-\lambda_1)(\epsilon_1^2+\epsilon_3^2)\\
+O(\epsilon_i^4)
\end{array},
& \begin{array}{c}     \\
  +\frac{\lambda_4}{2} t_\mu(\epsilon_1^2+\epsilon_3^2  \\
 - \epsilon_2^2-\epsilon_4^2) \\
  +\lambda_1\epsilon_1\epsilon_4+\lambda_2\epsilon_2\epsilon_3
\end{array}\\[0.5em]
\begin{array}{c} -\lambda_1( c_M\epsilon_4+ s_M\epsilon_2)\\
  +\lambda_4(- s_\mu\epsilon_1+  c_\mu\epsilon_4)\\
  +O(\epsilon_i^3)
\end{array},
& \begin{array}{c} \lambda_1 (s_M\epsilon_4-c_M\epsilon_2 )\\
  +\lambda_4(- s_\mu\epsilon_3+  c_\mu\epsilon_2)\\
  +O(\epsilon_i^3)
\end{array},
& \begin{array}{c}     \\
  +\frac{\lambda_4}{2} t_\mu(\epsilon_1^2+\epsilon_3^2  \\
 - \epsilon_2^2-\epsilon_4^2) \\
  +\lambda_1\epsilon_1\epsilon_4+\lambda_2\epsilon_2\epsilon_3
\end{array} ,
&\begin{array}{c} \lambda_4 + \\
-(\lambda_4-\lambda_1)(\epsilon_2^2+\epsilon_4^2)\\
+O(\epsilon_i^4)
\end{array}
\end{array}\right).\label{eq:MatrixDet1}
}
}
\noindent where we used $m_{G_1}=m_{G2}$ so that the small off-diagonal terms of (12) and (34) elements are almost zero, \ie $\lambda_1\simeq\lambda_2,~  \lambda_3\simeq\lambda_4$.
Thus, up to order $\epsilon_i^2$, we have
{\tiny
\dis{
(-\lambda_1 c_M+\lambda_4  c_\mu)\epsilon_1+~~0~~  -\lambda_1 s_M\epsilon_3
   + \lambda_4  s_\mu\epsilon_4 \simeq  \frac{M_Z}{\sqrt2}s_W(- c_\beta +s_\beta),   \\
    -\lambda_4 s_\mu\epsilon_1-\lambda_1 s_M\epsilon_2+~~0~~ +(-\lambda_1 c_M+\lambda_4  c_\mu)\epsilon_4
   \simeq  -\frac{M_Z}{\sqrt2}s_W (c_\beta+s_\beta) , \\
\lambda_1 s_M\epsilon_1 + \lambda_4  s_\mu\epsilon_2 +(-\lambda_1 c_M+\lambda_4  c_\mu)\epsilon_3 +~~0~~
    \simeq  \frac{M_Z}{\sqrt2}c_W (c_\beta-s_\beta),  \\
 ~~0~~+(-\lambda_1 c_M+\lambda_4  c_\mu)\epsilon_2- \lambda_4  s_\mu\epsilon_3 +\lambda_1 s_M\epsilon_4 \simeq   \frac{M_Z}{\sqrt2}c_W (c_\beta+s_\beta).
}
}
Thus, the solutions for $\epsilon_i$ in the limit $|\lambda_4| \gg |\lambda_1|$ are
\dis{
&\epsilon_1\simeq -\frac{M_Z}{\sqrt2\lambda_4} s_W\left[c_\mu  (c_\beta-s_\beta) -s_\mu(c_\beta+s_\beta) \right],~
\epsilon_2\simeq  \frac{M_Z}{\sqrt2\lambda_4} c_W\left[c_\mu  (c_\beta+s_\beta) +s_\mu(c_\beta-s_\beta) \right],\\
&\epsilon_3\simeq  \frac{M_Z}{\sqrt2\lambda_4}c_W \left[c_\mu(c_\beta-s_\beta) -s_\mu(c_\beta+ s_\beta)\right] ,~
\epsilon_4\simeq -\frac{M_Z}{\sqrt2\lambda_4} s_W\left[  s_\mu  (c_\beta-s_\beta)+ c_\mu (c_\beta+s_\beta) \right].
}
For the large parameters $\theta_M$ and $\theta_\mu$, we set the (12) and (34) components of Eq. (\ref{eq:MatrixDet1}) zero, in the limit $\lambda_1/\lambda_4\simeq 0$,
\dis{
\left.\begin{array}{l}
  \epsilon_1\epsilon_3 +\epsilon_2\epsilon_4 \simeq 0\\
 \epsilon_1\epsilon_4 +\epsilon_2\epsilon_3 \simeq 0
\end{array} \right\} \to ~~\epsilon_1^2\simeq \epsilon_2^2,~~\epsilon_3^2\simeq \epsilon_4^2 \label{eq:LarAngMuM}
}
which leads to
\dis{
(c_\mu^2-t_W^2 s_\mu^2)(c_\beta+s_\beta)^2+(s_\mu^2-t_W^2 c_\mu^2)(c_\beta-s_\beta)^2
+2c_\mu s_\mu (1+ t_W^2)(c_\beta^2-s_\beta^2)\simeq 0.
}
We notice that this has a solution for $s_\mu$ which needs not be small.
Namely, the elements $(\tilde U)_{33,34,43,44}$ elements are not very small.
Now, the original unitary matrix becomes
\dis{
U_{\chi^0}  =\left(\begin{array}{cccc}
c_M-\frac{1}{2c_M}(\epsilon_1^2+\epsilon_4^2), & s_M,  & \epsilon_1,& \epsilon_4\\
-s_M, &  c_M-\frac{1}{2c_M}(\epsilon_2^2+\epsilon_3^2),  & \epsilon_3, & \epsilon_2\\
-\frac{\epsilon_1+\epsilon_4}{\sqrt2}, & -\frac{\epsilon_2+\epsilon_3}{\sqrt2} , &\frac{c_\mu-s_\mu}{\sqrt2} -\frac{\epsilon_1^2+\epsilon_3^2}{2\sqrt2 c_\mu},  &  \frac{c_\mu+s_\mu}{\sqrt2} -\frac{\epsilon_2^2+\epsilon_4^2}{2\sqrt2 c_\mu} \\
\frac{-\epsilon_1+\epsilon_4}{\sqrt2}, &\frac{\epsilon_2-\epsilon_3}{\sqrt2},
& \frac{c_\mu+ s_\mu}{\sqrt2} -\frac{\epsilon_1^2+\epsilon_3^2}{2\sqrt2 c_\mu},
& \frac{-c_\mu+s_\mu}{\sqrt2} +\frac{\epsilon_2^2+\epsilon_4^2}{2\sqrt2 c_\mu}
\end{array}\right).
}

The smuon mass matrix is
\dis{
M^2_{\tilde\mu}=\left(\begin{array}{cc}
m_L^2 +M_Z^2 c_{2\beta}(s_W^2-\frac12), & m_\mu(A_\mu +\mu\tan\beta) \\
m_\mu(A_\mu +\mu\tan\beta),& m_R^2-M_Z^2 c_{2\beta} s_W^2
\end{array}\right)
}
The eigenvalues of $M^2_{\tilde\mu}$ are
\dis{
M^2_{\tilde\mu_{1,2}}=\frac12\left(m_L^2+m_R^2- \frac12 M_Z^2 c_{2\beta}\mp \sqrt{ M^4_{LRZ}  +4m_\mu^2(A_\mu  +  \mu t_\beta)^2\,}
\right)
}
\end{widetext}
where
\dis{
M^2_{LRZ}= m_L^2-m_R^2 +  M_Z^2 c_{2\beta} (2 s_W^2-\frac12) =|M^2_{LRZ}|\epsilon_{LRZ}
}
where $\epsilon_{LRZ}=\pm 1$ for the positive and negative $M^2_{LRZ}$, respectively.
The mass eigenstates $\tilde\mu _{1,2}$ are the mixtures of the left- and right-smuon states $\tilde\mu _{1,2}=\sin\gamma_{1,2} P_L +\cos\gamma_{1,2} P_R$
where $P_L=(1 ,~ 0)^T,~P_R=( 0,~ 1)^T$.
The smuon mixing angles are
\dis{
\sin\gamma_{1,2}=\frac{\sqrt2 m_\mu(  A_\mu + \mu t_\beta)}{|M^2_{LRZ}| \sqrt{D_N} } , }
\dis{
\cos\gamma_{1,2}=\frac{ \epsilon_{LRZ} \pm \sqrt{1+4m_\mu^2(A_\mu + \mu t_\beta)^2/M^4_{LRZ}\,}  }{\sqrt2 \sqrt{ D_N} }\label{eq:sincos}
}
where
\dis{
D_N= 1\pm\sqrt{S}+\frac{4m_\mu^2(A_\mu + \mu t_\beta)^2}{M^4_{LRZ}},
}
with
\dis{
S=1 + \frac{4m_\mu^2(A_\mu + \mu t_\beta)^2}{M^4_{LRZ}}.
}

So,
\dis{
 N^L_{AX}&=-y_\mu (U_{\chi^0})_{3X}(U_{\tilde\mu})_{LA}-\sqrt2 g_1 (U_{\chi^0})_{1X}(U_{\tilde\mu})_{RA}\\
 &\simeq -\sqrt2 g_1 (U_{\chi^0})_{1X}\cos\gamma_{1,2}\\
 N^R_{AX}&=-y_\mu (U_{\chi^0})_{3X}(U_{\tilde\mu})_{RA}-\frac{1}{\sqrt2} g_2 (U_{\chi^0})_{2X}(U_{\tilde\mu})_{LA}\\
 &~~-\frac{1}{\sqrt2} g_1 (U_{\chi^0})_{1X}(U_{\tilde\mu})_{LA}\\
 &\simeq -\frac{1}{\sqrt2} g_2 (U_{\chi^0})_{2X}\sin\gamma_{1,2}-\frac{1}{\sqrt2} g_1 (U_{\chi^0})_{1X}\sin\gamma_{1,2}
}
Therefore,
\dis{
\sum_X &N^L_{AX}N^R_{AX}\simeq\sum_X   [g_1 g_2 (U_{\chi^0})_{1X}  (U_{\chi^0})_{2X}\\
 &~~~+ g_1^2 (U_{\chi^0})_{1X} (U_{\chi^0})_{1X}  ] \cos\gamma_{1,2} \sin\gamma_{1,2}}
 \dis{
& = \cos\gamma_{1,2} \sin\gamma_{1,2} \Big\{ g_1 g_2[(U_{\chi^0})_{11}  (U_{\chi^0})_{21} \\
&~~~+(U_{\chi^0})_{12} (U_{\chi^0})_{22}] +g_1^2 [(U_{\chi^0})_{11} (U_{\chi^0})_{11}\\
&~~~+(U_{\chi^0})_{12} (U_{\chi^0})_{12}]\Big\}\nonumber
}
\dis{
&\to  \cos\gamma_{1,2} \sin\gamma_{1,2}\Big[g_1 g_2\frac{s_M}{2c_M}(\epsilon_1^2-\epsilon_2^2-\epsilon_3^2+\epsilon_4^2 )\\
&~~~~~~~~~~~~+g_1^2(1-\epsilon_1^2-\epsilon_4^2) \Big]
}
\dis{
\to g_1^2\cos\gamma_{1,2} \sin\gamma_{1,2}.\nonumber
}
Numerically, this becomes
\dis{
G^2 s_W^2 &\cos\gamma_{1,2}\sin\gamma_{1,2}
 =\frac{e^2}{c_W^2}\cos\gamma_{1,2}\sin\gamma_{1,2}\\
 &\simeq~~~ 0.12 \cos\gamma_{1,2}\sin\gamma_{1,2}.\label{eq:sumI}
}
Note that the U(1)$_Y$ gauge contribution dominates. So, we estimate for $\epsilon_{LRZ}=1$
\dis{
&\frac{a_\mu^{\rm SUSY}(\chi^{0A})}{1\times 10^{-9}}  \simeq -1.32\, t_\beta \sum_{i=\rm lighter~one}\, \left(\frac{\mu}{10\,\tev} \right)\\[0.3em]
&~~~\cdot\left(\frac{(300\,\gev)^2}{|M^2_{LRZ}|} \right)
\left(\frac{\tev}{\tilde\mu_{i}} \right)^2 \left(\frac{m_{\chi^{0A}_I}}{\tev} \right)\cdot I^0(x_{Ai})\\ &~
}
where $x_{Ai}=m^2_{\chi^0_I}/m^2_{\tilde \mu_i}$, and $\chi^{0A}_I$ corresponds to two small eigenvalues of the $2\times 2$ matrix elements of the upper left corner of $U_{\chi^0}$. Since $\chi^{0A}_1$ and $\chi^{0A}_2$ are almost degenerate, the sum over $I$ is aleady taken into account in Eq. (\ref{eq:sumI}). The sum over $i$ is dominated by the lighter smuon and the coefficient --1.32 corresponds to the lighter smuon. For the heavier smuon, the coefficient is +1.86. Since $t_\beta$ is of order 10, the neutralino contribution gives a correct order of the muon anomalous magnetic moment observed at BNL. To have the positive sign for $a_\mu^{\rm SUSY}(\chi^{0A})$, the product of signs of $t_\beta,\mu, m_{\chi^{0A}_I}$, and $I^0(x_{Ai})$ should be negative where
\dis{
I^0(x)=\frac{1-x^2+2x\ln x}{(1-x)^3}.
}

\vskip 0.2cm

\noindent {\bf Acknowledgments}: {I would like to thank Min-Seok Seo for useful discussions and especially Ji-Haeng Huh for assisting in drawing the scatter plots.
This work is supported in part by the National Research Foundation (NRF) grant funded by the Korean Government (MEST) (No. 2005-0093841).
}

\vskip 1cm

%%%%%%%%%%%%%%%%%%%%%%%%%%%%%%%%%%%%%%%%%%%%%%%%%%%%%%%%%%%%%%%%%%%%%%%%%%%%%%%%%%
%%%%%%%%%%%%%%%%%%%%%%%%%%%%%%%%%%%%%%%%%%%%%%%%%%%%%%%%%%%%%%%%%%%%%%%%%%%%%%%%%%
%%%%%%%%%%%%%%%%%%%%%%%%%%%%%%%%%%%%%%%%%%%%%%%%%%%%%%%%%%%%%%%%%%%%%%%%%%%%%%%%%%

%\newpage


\begin{thebibliography}{99}

\def\prp#1#2#3{Phys.\ Rep.\ {\bf #1} (#3) #2}
\def\rmp#1#2#3{Rev. Mod. Phys.\ {\bf #1} (#3) #2}
\def\npb#1#2#3{Nucl.\ Phys.\ {\bf B#1} (#3) #2}
\def\plb#1#2#3{Phys.\ Lett.\ {\bf B#1} (#3) #2}
\def\prd#1#2#3{Phys.\ Rev.\ {\bf D#1} (#3) #2}
\def\prl#1#2#3{Phys.\ Rev.\ Lett.\ {\bf #1} (#3) #2}
\def\err#1#2#3{\ {\bf #1} (#3) #2\,(E)}
\def\jhep#1#2#3{JHEP\ {\bf #1} (#3) #2}
\def\jcap#1#2#3{JCAP\ {\bf #1} (#3) #2}
\def\zp#1#2#3{Z.\ Phys.\ {\bf #1} (#3) #2}
\def\epjc#1#2#3{Euro. Phys. J.\ {\bf C#1} (#3) #2}
\def\ijmp#1#2#3{Int.\ J.\ Mod.\ Phys.\ {\bf #1} (#3) #2}
\def\mpl#1#2#3{Mod.\ Phys.\ Lett.\ {\bf A#1} (#3) #2 }
\def\apj#1#2#3{Astrophys.\ J.\ {\bf #1} (#3) #2}
\def\nat#1#2#3{Nature\ {\bf #1} (#3) #2}
\def\sjnp#1#2#3{Sov.\ J.\ Nucl.\ Phys.\ {\bf #1} (#3) #2}
\def\apj#1#2#3{Astrophys.\ J.\ {\bf #1} (#3) #2}
\def\ijmp#1#2#3{Int.\ J.\ Mod.\ Phys.\ {\bf #1} (#3) #2}
\def\mpl#1#2#3{Mod.\ Phys.\ Lett.\ {\bf A#1} (#3) #2 }
\def\nat#1#2#3{Nature\ {\bf #1} (#3) #2}
\def\npb#1#2#3{Nucl.\ Phys.\ {\bf B#1} (#3) #2}
\def\pthp#1#2#3{Prog.\ Theor.\ Phys.\ {\bf #1} (#3) #2}


\bibitem{CERNJuly4} The CMS Collaboration, `\emph{Observation of a new boson at a mass of 125 GeV with the CMS experiment at the LHC}' [arXiv:1207.7235];  The ATLAS Collaboration, `\emph{Observation of a New Particle in the Search for the Standard Model Higgs Boson with the ATLAS Detector at the LHC}' [arXiv:1207.7214].

\bibitem{LHC3rd} ATLAS Collaboration, `\emph{Search for scalar top quark pair production in natural gauge mediated supersymmetry models with the ATLAS detector in pp collisions at $\sqrt{s} = 7$ TeV},' [arXiv:1204.6736];  `\emph{Hunt for new phenomena using large jet multiplicities and missing transverse momentum with ATLAS in 4.7 ${\rm fb}^{-1}$ of $\sqrt{s} = 7$ TeV proton-proton collisions},'  [arXiv:1206.1760].

\bibitem{LHCsquark} CMS Collaboration, `\emph{Search for supersymmetry at the LHC in events with jets and missing transverse energy},' \prl{107}{221804}{2011} [arXiv:1109.2352].

\bibitem{effSUSY95} S. Dimopoulos and G. G. Giudice, `\emph{Naturalness constraints in supersymmetric theories with non-universal soft terms},' \plb{357}{573}{1995}   [hep-ph/9507282]; A Pomarol and Tommasini, `\emph{Horizontal symmetries for the supersymmetric flavor problem},' \npb{466}{3}{1995}  [hep-ph/9507462];
    A. Cohen, D. B. Kaplan, and A. E. Nelson, `\emph{The more minimal supersymmetric standard model},' \plb{388}{588}{1996} [hep-ph/9607394]; G. G. Giudice-Nardecchia-Romanino, `\emph{Hierarchical soft terms and flavor physics},' \npb{813}{156}{2009} [arXiv:0812.3610 [hep-ph] ].

\bibitem{HiggsBound}
H.E. Haber and R. Hempfling, `\emph{Can the mass of the lightest Higgs boson of the minimal supersymmetric model be larger than m(Z)?},' \prl{66}{1815}{1991};
Y. Okada, M. Yamaguchi and T. Yanagida, `\emph{Upper bound of the lightest Higgs boson mass in the minimal supersymmetric standard model},' \pthp{85}{1}{1991};
J. Ellis, G. Ridolfi and F. Zwirner, `\emph{Radiative corrections to the masses of supersymmetric Higgs bosons},' \plb{257}{83}{1991}.

%\bibitem{LargeA} One can also raise by introducing more fields: M. Asano, T. Moroi, R. Sato, and T. T. Yanagida, \plb{705}{337}{2011}; J. L. Evans, M. Ibe, and T. T. Yanagida, [arXiv:1108.3437]; T. Moroi, R. Sato, and T. T. Yanagida, \plb{709}{218}{2012}.

\bibitem{Jeong11} K. S. Jeong, J. E. Kim, and M. S. Seo, `\emph{Gauge mediation to effective SUSY through U(1)s with a dynamical SUSY breaking, and string compactification},' \prd{84}{075008}{2011} [arXiv:1107.5613].

\bibitem{Lang08} P. Langacker, G. Paz, L.-T. Wang, and I. Yavin, `\emph{$Z'$-mediated supersymmetry breaking},' \prl{100}{041802}{2008} [arXiv:0710.1632 [hep-ph]].

\bibitem{Zpfamind}  R. N. Mohapatra and S. Nandi, `\emph{A new messenger sector for gauge mediated supersymmetry breaking},' \prl{79}{181}{1997} [hep-ph/9702291]; T. Kikuchi and T. Kubo, `\emph{Radiative B-L symmetry breaking and the $Z'$ mediated SUSY breaking},' \plb{666}{262}{2008} [arXiv:0804.3933 [hep-ph]]; AIP Conf. Proc. {\bf 1078} (2009) 402 [arXiv:0809.2011 [hep-ph]].

\bibitem{ArkaMura97} N. Arkani-Hamed and H. Murayama, `\emph{Can the supersymmetric flavor problem decouple?},' \prd{56}{6733}{1997},[arXiv:hep-ph/9703259]

%\bibitem{Tamarit12} C. Tamarit, `\emph{Large, negative threshold contributions to
% light soft masses in effective SUSY scenarios}' [arXiv:1206.6140], and references therein.

\bibitem{Hisano99}  J. Hisano, K. Kurasawa, and Y. Nomura, `\emph{Large squark and slepton masses for the first-two generations in the anomalous U(1) SUSY breaking models},' \plb{445}{316}{1999} [arXiv:hep-ph/9810411].

\bibitem{Nilles84} H. P. Nilles, `\emph{Supersymmetry, supergravity and particle physics},' \prp{110}{1}{1984}, and `\emph{Dynamically broken supergravity and the hierarchy problem},' \plb{115}{193}{1982}.

\bibitem{gmtwoBNL} G. W. Bennett \etal, `\emph{Measurement of the positive muon anomalous magnetic moment to 0.7 ppm},' \prl{89}{101804}{2002}, Erratum \prl{89}{129903}{2002};
    G. W. Bennett  \etal, `\emph{Measurement of the negative muon anomalous magnetic moment to 0.7 ppm},' \prl{92}{161802}{2004}; Phys. Rev. Lett. 92, 161802 (2004);
    G. W. Bennett \etal, `\emph{Final report of the muon E821 anomalous magnetic moment measurement at BNL},' \prd{73}{072003}{2006}.

\bibitem{MixedMed2} G. Hiller, Y. Hochberg, and Y. Nir, `\emph{Flavor changing processes in supersymmetric models with hybrid gauge- and gravity-mediation},' \jhep{0903}{115}{2009} [arXiv:0812.0511].

\bibitem{MixedMed3} L. L. Everett, I.-W. Kim, P. Ouyang, and K. M. Zurek, `\emph{Deflected mirage mediation: A framework for generalized supersymmetry breaking},' \prl{101}{101803}{2008}  [arXiv:0804.0592], and `\emph{Moduli stabilization and supersymmetry breaking in deflected mirage mediation},' \jhep{0808}{102}{2008} [arXiv:0806.2330].

\bibitem{Kim07stable} J. E. Kim, `\emph{GMSB at a stable vacuum and MSSM without exotics from heterotic string,}'  \plb{656}{207}{2007}  [arXiv: 0707.3292[hep-ph]].

\bibitem{HuhKK09} J.-H. Huh, J. E. Kim and B. Kyae, `\emph{SU(5)$_{flip}\times$SU(5)$'$ from {\bf Z}$_{12-I}$},' \prd{80}{115012}{2009} [arXiv:0904.1108].

\bibitem{SplitSUSY}  N. Arkani-Hamed and S. Dimopoulos, `\emph{Supersymmetric unification without low energy supersymmetry and signatures for fine-tuning at the LHC},' \jhep{0506}{073}{2005}  [hep-th/0405159];
     G. Giudice and A. Romanino, `\emph{Split supersymmetry},' \npb{699}{65}{2004} [hep-ph/0406088];
     N. Arkani-Hamed, S. Dimopoulos, G. Giudice, and A. Romanino, `\emph{Aspects of split supersymmetry},' \npb{709}{3}{2005} [hep-ph/0409232].

\bibitem{Chua12} C. K. Chua, `\emph{Implications of $B(\mu\to e\gamma)$ and $\Delta a_\mu$ on muonic lepton flavor violating processes},' [arXiv:1205.3898 [hep-ph]].

\bibitem{KMO74} J. E. Kim, V. S. Mathur, and S. Okubo, `\emph{Electromagnetic properties of  the  neutrino  from neutral current experiments},' \prd{9}{3050}{1974}. For the WIMP magnetic moment, see,  W. S. Cho, J.-H. Huh, J. E. Kim, I.-W. Kim, and B. Kyae, `\emph{Constraining WIMP magnetic moment from CDMS II experiment},' \plb{687}{6}{2010}, and \err{B694}{496}{2011} [arXiv: 1001.0579[hep-ph]].

\bibitem{Kim76} J. E. Kim, `\emph{Neutrino  magnetic moment},' \prd{14}{3000}{1976}.

\bibitem{Moroi96} T. Moroi, `\emph{The muon anomalous magnetic dipole moment in the minimal supersymmetric standard model},' \prd{53}{6565}{1996}  [arXiv:hep-ph/9512396].

\bibitem{Adam11} J. Adam \etal, `\emph{New limit on the lepton-flavour violating decay $\mu\to e\gamma$},' [MEG Collaboration], \prl{107}{171801}{2011}  [arXiv:1107.5547[hep-ex]].

\bibitem{Masiero07}  A. Masiero, S. K. Vempati, and O. Vives, `\emph{Flavour physics and grand unification,}' [arXiv:0711.2903].

\bibitem{SOFTSUSY12} B. C. Allanach, `\emph{SOFTSUSY: a program for calculating supersymmetric spectra},' arXiv:hep-ph/0104145v17 (2012).

\bibitem{bsgamma} F. M. Borozumati and C. Greub, `\emph{Two Higgs doublet model predictions for $\overline{B}\to X_s\gamma$ in NLO QCD},' \prd{58}{074004}{1998} [hep-ph/9802391]; M. Neubert, `\emph{Renormalization-group improved calculation
of the ${B}\to X_s\gamma$ branching ratio},' \epjc{40}{165}{2005} [hep-ph/0408179].

\bibitem{AxinoToLSP}  K.-Y. Choi, J. E. Kim, H. M. Lee, and O. Seto, `\emph{Neutralino dark matter from heavy axino decay,}' \prd{77}{123501}{2008} [arXiv: 0801.0491[hep-ph]]; J. E. Kim and M.-S. Seo, `\emph{Mixing of axino and goldstino, and axino mass,}' \npb{864}{296}{2012}  [arXiv:1204.5495 [hep-ph]].

\end{thebibliography}
\end{document}